# Adjustable Nonlinear Springs to Improve Efficiency of Vibration Energy Harvesters

SEBASTIEN BOISSEAU, GHISLAIN DESPESSE AND BOUHADJAR AHMED SEDDIK
*CEA-LETI, Minatec campus, 17, av. des martyrs, 38000 Grenoble, France*

**ABSTRACT:** Vibration Energy Harvesting is an emerging technology aimed at turning mechanical energy from vibrations into electricity to power microsystems of the future. Most of present vibration energy harvesters are based on a mass spring structure introducing a resonance phenomenon that allows to increase the output power compared to non-resonant systems, but limits the working frequency bandwidth. Therefore, they are not able to harvest energy when ambient vibrations' frequencies shift. To follow shifts of ambient vibration frequencies and to increase the frequency band where energy can be harvested, one solution consists in using nonlinear springs. We present in this paper a model of adjustable nonlinear springs (H-shaped springs) and their benefits to improve velocity-damped vibration energy harvesters' (VEH) output powers. A simulation on a real vibration source proves that the output power can be higher in nonlinear devices compared to linear systems (up to +48%).

## INTRODUCTION

MEMS and smart material technologies improvements have allowed autonomous sensor devices to become more and more widespread over the past few years. As batteries are not always appropriated to power these systems, energy harvesting from ambient environment solutions are currently being developed. Among the potential energy sources, we have focused on mechanical surrounding vibrations. Thanks to measurements and in agreement with recent studies (Roundy, 2003; Despesse, 2005), we have observed that most of surrounding mechanical vibrations occur at frequencies below 100 Hz, can be spread on a large frequency bandwidth (1 to 100 Hz) and can move through time. Vibration energy harvesters (VEH) are generally based on a mass-spring structure bringing a phenomenon of resonance (Anton and Sodano, 2007; Beeby, Tudor and White, 2006; Cook-Chennault, Thambi and Sastry, 2008; Saadon and Sidek, 2011). Unfortunately, the phenomenon of resonance has a tight working frequency bandwidth. As a consequence, VEH cannot extract energy when ambient vibrations' frequencies shift (e.g. car engine).

A way to follow ambient vibration frequencies' shifts is to take advantage of nonlinear behaviors. These nonlinear effects and these behaviors can be introduced in VEH by the use of nonlinear springs or non-linear stiffnesses (Amri et al., 2011; Ando et al., 2010; Cottone, Vocca and Gammaitoni, 2009; Ferrari et al., 2009; Quinn et al., 2011; Mann and Owens, 2010; Miki et al., 2010; Nguyen and Halvorsen, 2010; Stanton and Mann, 2010; Tang, Yang and Soh, 2010; Triplett and Quinn, 2009). While the research on nonlinear effects in VEH is quite recent, nonlinear springs' behaviors are well known for a long time and many models have already been developed (Kaajakari et al., 2004; Landau and Lifshitz, 1999; Legtenberg, Groeneveld and Elwenspoek, 1996; Marzencki, Defosseux and Basrour, 2009). They are based on the fact that, contrary to linear springs where the spring force is proportional to the deformation, the spring force of nonlinear springs is a polynomial function (with a degree higher than 1) of the deformation.

In this paper, we focus on the study of clamped-guided beam which is one of the most common springs in MEMS (Legtenberg, Groeneveld and Elwenspoek, 1996; Lobontiu and Garcia, 2005). After developing velocity-damped VEH' generic linear model, we focus on clamped-guided beams and their effects on VEH behavior. Our mechanical equations are validated by finite element analyses (FEA). Then, we present a new design for clamped-guided beam allowing adjusting and optimizing VEH response. We finish this study by applying the adjusted nonlinear effects on a real vibration source, proving the interest of nonlinear VEH in specific environments. Finally a nonlinear mechanical structure, optimized for a specific vibration source, is sized.

## LINEAR MODEL OF VELOCITY-DAMPED VIBRATION ENERGY HARVESTERS

Whatever the principle of conversion used (piezoelectric, electromagnetic or electrostatic), resonant VEH can be represented as a mobile mass (*m*) attached to a frame by a spring (*k*) and damped by mechanical and electrical forces ($f_{mec}$ and $f_{elec}$). When the system is subjected to ambient vibrations $a_v(t)$, a relative displacement *x(t)* between the mobile mass and the frame appears (figure 1). Part of the mobile mass's kinetic energy is lost in mechanical damping ($f_{mec}$), which is generally modeled as a linear viscous force, while the other part is turned into electricity thanks to an energy converter (piezoelectric material, magnet/coil, variable capacitor,…), and modeled as an electric force ($f_{elec}$). Ambient vibrations are generally low in amplitude (~25-50µm) and in frequencies (<100Hz); the use of a mass-spring structure allows to amplify the vibrations perceived by the mobile mass thanks to the resonance and to increase VEH output power.

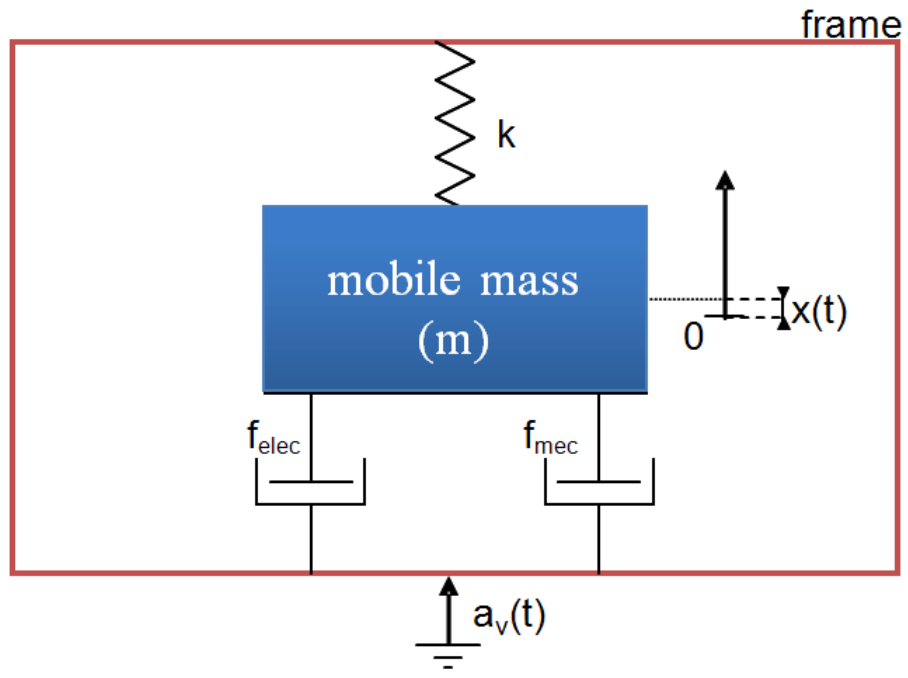

**Figure 1. Generic model of energy harvesters**

The fundamental principle of dynamics gives a simple differential equation that rules the mobile mass's movement (1).

$$m\ddot{x} + kx + f_{elec} + f_{mec} = -ma_v \quad (1)$$

However, the equation that rules $f_{elec}$ may be complicated, generally linked to coupled differential equations and especially in the case of piezoelectric (Erturkand Inman, 2008) and electrostatic (Boisseau, Despesse and Sylvestre, 2010) devices. In the case of electromagnetic devices that can be considered as velocity damped VEH, $f_{elec}$ and equation (1) can be greatly simplified.

Actually, in a model developed by William and Yates (1996) for electromagnetic devices, $f_{mec}$ and $f_{elec}$ are represented as linear viscous forces, $f_{elec} = b_e\dot{x}$ and $f_{mec} = b_m\dot{x}$, where $b_e$ and $b_m$ are respectively electrical and mechanical damping coefficients.

In this paper, as our objective is to show the interest of nonlinear H-shaped springs to increase VEH output power, we will focus on the simpler model, i.e. William & Yates's model for velocity-damped devices.

From William & Yates's model, equation (1) can be simplified by using VEH angular natural frequency ($\omega_0 = \sqrt{k/m}$) and electrical and mechanical damping rates, $\xi_e = b_e/(2m\omega_0)$ and $\xi_m = b_m/(2m\omega_0)$.

$$\ddot{x} + 2(\xi_m + \xi_e)\omega_0\dot{x} + \omega_0^2 x = -a_v \tag{2}$$

$\xi_m$ is often replaced by the mechanical quality factor *Q*, and the link between them is $\xi_m = 1/2Q$.

For a mono-frequency vibration $a_v(t) = A\sin(\omega t)$, velocity-damped VEH maximum output power is reached when its natural frequency ($f_0 = \frac{1}{2\pi}\omega_0$) is tuned to the ambient vibrations frequency ($f = \frac{1}{2\pi}\omega$) and when the damping rates $\xi_e$ and $\xi_m$ are equal. This maximum output power $P_{W\&Y}$ can be simply expressed with (3).

$$P_{W\&Y} = \frac{mA^2Q}{8\omega_0} \tag{3}$$

This model is in agreement with experiments until the spring leaves its linear behavior and leads to nonlinear effects. They become significant when the working displacements become higher than the beam thickness and especially for clamped-clamped or clamped-guided beams.

In the next section, we develop a model of a clamped-guided beam taking into account nonlinear behaviors deduced from the theory of energy conservation. The results are validated by FEA.

**CLAMPED-GUIDED BEAM AND NONLINEAR EFFECTS**

The clamped-guided beam studied in this paper is presented in figure 2, with *L* its length, *b* its width and *e* its thickness. The beam is clamped on its left and guided on its right, meaning that only the displacement in $\vec{e}_x$ direction is allowed. To determine the equivalent spring constant *k*, the ratio between the displacements at the right end of the beam and the force *F* applied at this same right end is computed.

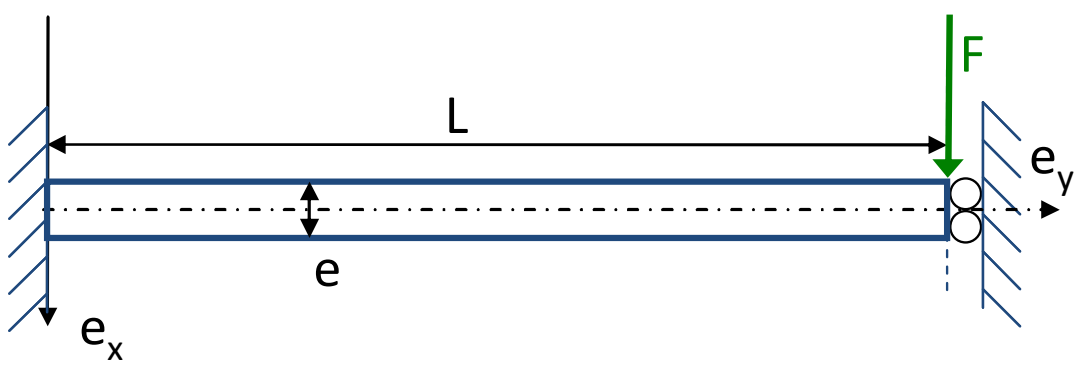

**Figure 2. Clamped-guided beam**

In the linear case, this ratio does not depend on the value of *F*; but by introducing nonlinear mechanical effects that we look for, this ratio becomes non-constant. To present the differences between linear and nonlinear behaviors, we study both cases in the following sub-parts.

**Linear behavior**

From the theory of structural mechanics, one can find the bending moment along $\vec{e}_y$ axis *M(y)* of the clamped-guided beam presented in figure 2.

$$M(y) = -F\left(y - \frac{L}{2}\right) \tag{4}$$

As $x(y) = \iint \frac{M(y)}{EI} dy$, where *E* is the Young's Modulus of the material and *I* the beam second moment of area, it is possible to determine the deflection *x* on each point of the beam and therefore the linear equivalent spring stiffness $k_0$ in the case of small displacements.

$$x(y) = \frac{F}{EI}\left(\frac{y^3}{6} - \frac{Ly^2}{4}\right) \Rightarrow |x(L)| = \frac{FL^3}{12EI} \Rightarrow k_0 = \frac{F}{|x(L)|} = \frac{12EI}{L^3} = \frac{Ebe^3}{L^3} \tag{5}$$

This well-known formula is only valid if nonlinear effects are neglected.

**Nonlinear effects – Beam elongation contribution**

Nonlinear effects are well-known in springs when the relative displacement is high compared to the beam thickness in the displacement direction. We can separate springs into 3 categories: linear springs, hardening springs and softening springs depending on their behavior with regard to the deformation *x* (figure 3).

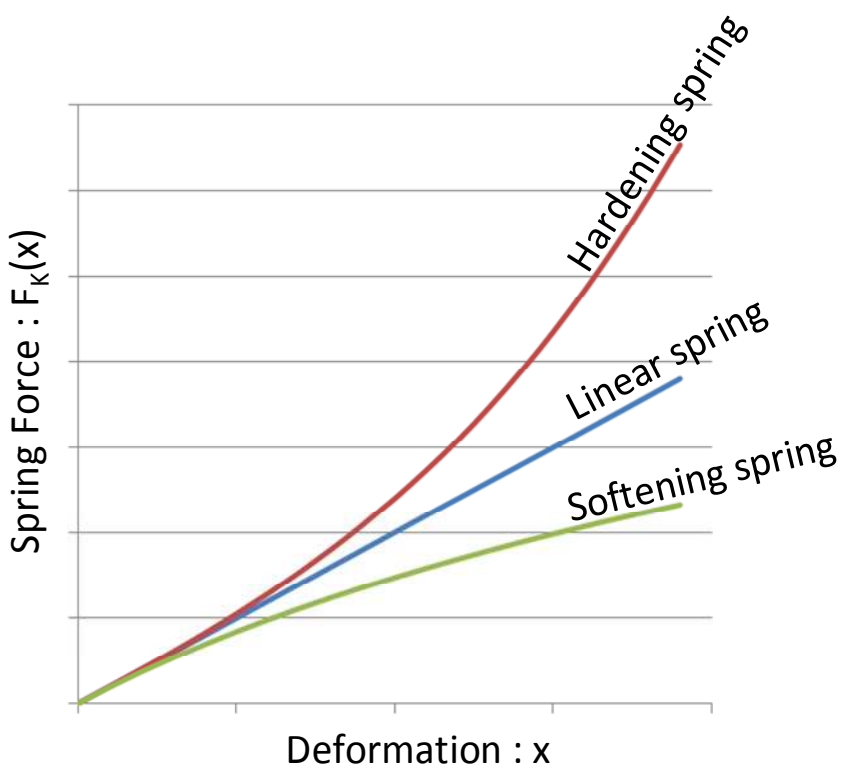

**Figure 3. Characterization of springs**

Nonlinear springs' behavior is often represented by a deformation-dependant spring constant *k(x)*, a function of the linear spring constant $k_0$ and a non-constant part $k_{nl}(x)$:

$$F_K(x) = k_0 x + f(x) \Rightarrow k(x) = \frac{F_K(x)}{x} = \frac{k_0 x + f(x)}{x} = k_0 + k_{nl}(x) \tag{6}$$

In clamped-guided beams, nonlinear behaviors are mainly due to traction effects, which are not taken into account in the linear model. To calculate the contribution of the elongation force in reaction to the *F* force, we use the theory of energy conservation: we consider that the energy of the bending force is entirely converted into elongation due to traction force (7). The elongation *δL* of the beam length due to the force *F* is presented in figure 4.

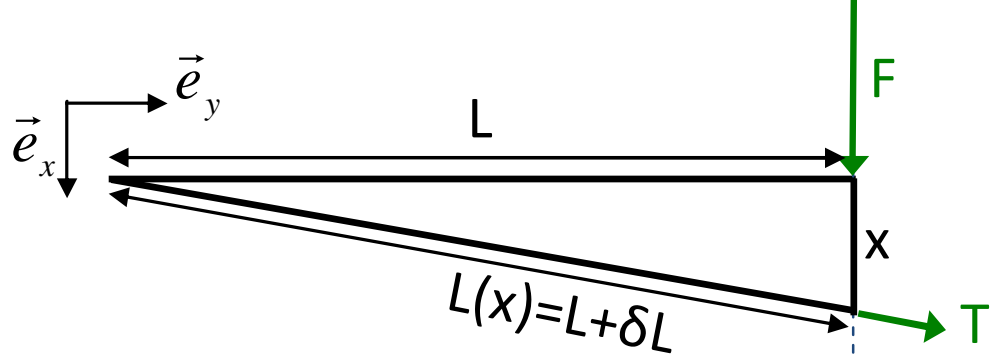

**Figure 4. Nonlinear effects in clamped-guided beams**

By applying the energy conservation principle and considering only the energy part stored in beam elongation, the mechanical work of the force *F* on the infinitesimal displacement *dx* is equal to the mechanical work associated to the infinitesimal elongation *dL(x)* of the beam.

$$Fdx = TdL(x) \tag{7}$$

By applying Pythagorean Theorem to figure 4, we can deduce the beam elongation $\delta L$ induced by a displacement *x* of the beam guided end in $\vec{e}_x$ direction:

$$L^2 + x^2 = (L+\delta L)^2 \Rightarrow \delta L = \sqrt{L^2+x^2} - L \tag{8}$$

As $L(x) = L + \delta L$, and since *L* is a (constant) parameter, we get $dL(x) = d(\delta L)$ that is simplified into $dL = d(\delta L)$.

Then, by a derivation of equation (8):

$$\frac{dL}{dx} = \frac{d(\delta L)}{dx} = \frac{x}{L\sqrt{1+\left(\frac{x}{L}\right)^2}} \tag{9}$$

The beam elongation constraint *T* can be expressed in function of its elongation ratio by using Hooke's law: $T = \left(\frac{dL}{L}ES\right)$ with *S* the beam section *S=eb.*

Finally, the relation *F(x)* between the applied force *F* and the induced displacement *x* can be deduced from equation (7) and by replacing some parts by the previous relations (equations (8) and (9)).

$$F(x) = T\left(\frac{dL}{dx}\right) = \left(\frac{dL}{L}ES\right)\left(\frac{dL}{dx}\right) = \left(\sqrt{1+\left(\frac{x}{L}\right)^2} - 1\right)ES\frac{x}{L\sqrt{1+\left(\frac{x}{L}\right)^2}} \tag{10}$$

Thanks to a Taylor expansion to order 2:

$$F(x) = \underbrace{\left(\frac{ES}{L}\right)}_{k_{trac}} x\left(\frac{1}{2}\left(\frac{x}{L}\right)^2\right) \Rightarrow F(x) = k_0 x \frac{1}{2}\left(\frac{x}{e}\right)^2 \tag{11}$$

It is important to notice that $k_{trac}$, the beam stiffness in traction appears in the third order spring force (11). Finally, by taking both effects of deflection (linear model) and traction (beam elongation contribution), one can find that the spring force ($F_k$) is equal to:

$$F_k(x) = k_0 x\left(1+\frac{1}{2}\left(\frac{x}{e}\right)^2\right) \tag{12}$$

This expression is in agreement with other theories (and experiments) that model the spring force by $F_k(x) = k_1 x + k_3 x^3$ (Duffing, 1918; Marzencki, Defosseux and Basrour, 2009; Wiggins, 1990), an odd polynomial function *K[x]* of degree 3. We have chosen to limit our study to the order 3, but, a study at a superior order could be possible (Nguyen and Halvorsen, 2010). In the following part we check this formula by FEA.

**Validation with finite element analyses**

To check our analytical model of the spring-force relation with FEA, we have chosen Comsol® Multiphysics, which allows to take nonlinear effects of structural mechanics into account.

FINITE ELEMENT (FEA) MODEL

In Comsol® Multiphysics, in order to see the effects of nonlinear behaviors, it is necessary to turn the “large deformations” option *on* (available in particular in MEMS module). Therefore, we have chosen the module “MEMS/Structural Mechanics/Plane Stress/Static Analysis”. So as to compare our analytical model to FEA, we have chosen to work on a silicon beam (*E*=131GPa), 1cm long (*L*), 1mm wide (*b*) and 100µm thick (*e*).

Boundary conditions are deduced from figure 5: the left boundary is fixed, the right boundary has an imposed displacement *y=0*; the two other boundaries are left free.

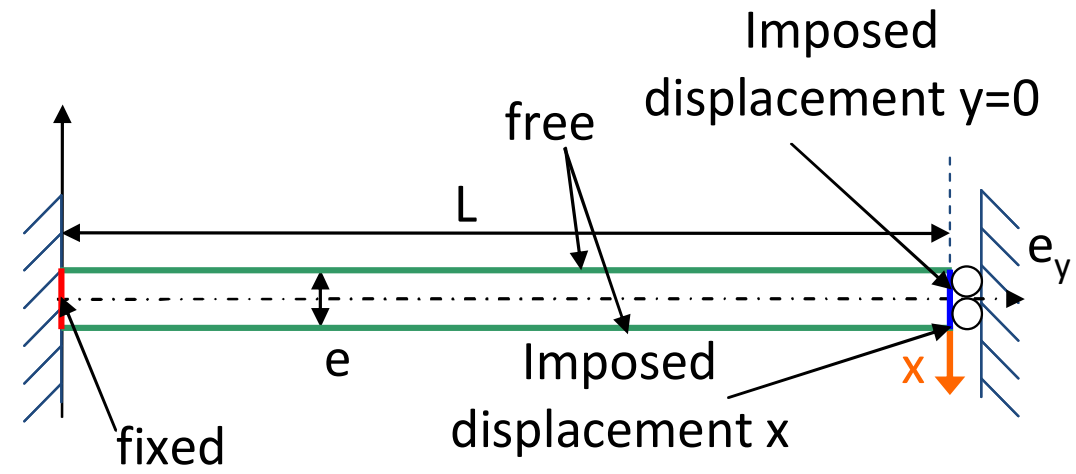

**Figure 5. Boundary conditions for FEA model**

An imposed displacement *x* is applied on the lower part of the right end of the beam and the associated reaction force is extracted from Comsol®. This spring force-displacement relation is presented in the following part and compared to the analytical model.

ANALYTICAL NONLINEAR BEAM MODEL VS FEA RESULTS

To compare our analytical expression of the spring force-displacement relation to FEA, we compute the derivatives of the spring force ($F_k$) with respect to *x* which are easy to extract from Comsol® software:

$$\begin{cases} F_K(x) = k_0 x + k_3 x^3 + o(x^4) \\ \dfrac{dF_K}{dx}(x) = k_0 + 3k_3 x^2 + o\left(x^3\right) \\ \dfrac{d^2 F_K}{dx^2}(x) = 6k_3 x + o\left(x^2\right) \\ \dfrac{d^3 F_K}{dx^3}(x) = 6k_3 + o\left(x\right) \end{cases} \tag{13}$$

Therefore, from $\dfrac{dF_K}{dx}(x)$ and $\dfrac{d^3 F_K}{dx^3}(x)$, it is possible to find the values of the linear spring constant $k_0$ and the third order spring constant $k_3$. Figure 6 presents the spring force and its successive derivatives with respect to the displacement *x*.

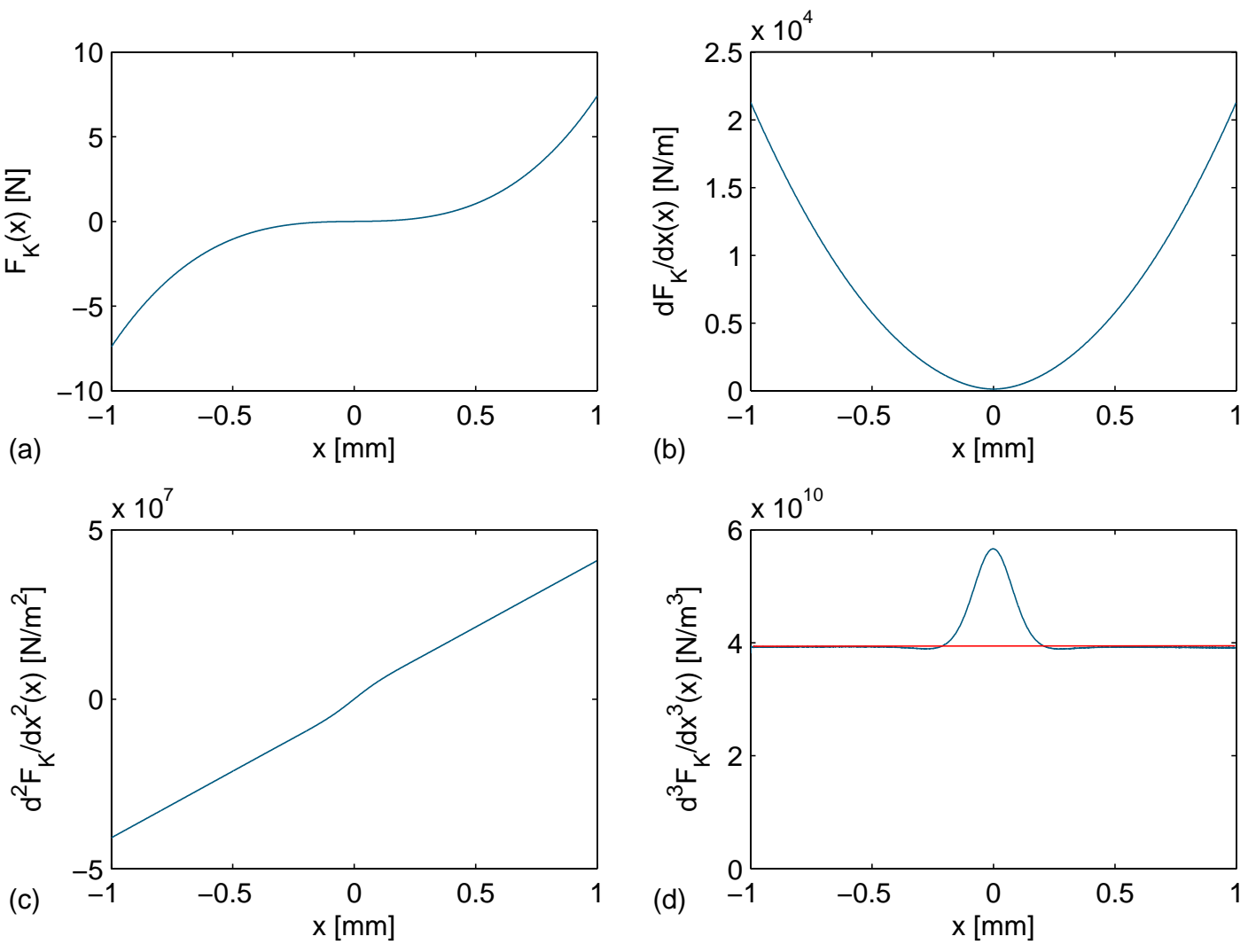

**Figure 6. Successive derivatives of $F_K(x)$ with respect to x: (a) $F_K$, (b)$F'_K$ (c)$F''_K$, (d)$F'''_K$**

FEA results are in agreement with our model except $\frac{d^3F_K}{dx^3}(x)$ as it should be a constant (figure 6d). This difference around *x*=0 is due to calculus errors of the third derivative of $F_K(x)$ from $F_K(x)$ provided by FEA. Nevertheless, by moving away from *x=0*, the value of $\frac{d^3F_K}{dx^3}(x)$ is constant and close to $6k_3$; this is verified in figure 6d.

Table 1 compares analytical and FEA values of $k_0$ and $k_3$. We can note that our theoretical study is in agreement with FEA as these two values only differ by 0.25% in the worst case.

**Table 1. Theoretical and FEA values of $k_0$ and $k_3$**

| Value | Theory | Theoretical value | FEA value | Difference |
|---|---|---|---|---|
| $k_0$ | $k_0 = \frac{Ebe^3}{L^3}$ | 131 $N.m^{-1}$ | 131 $N.m^{-1}$ | 0% |
| $k_3$ | $k_3 = \frac{k_0}{2e^2}$ | $6.55\times10^9$ $N.m^{-3}$ | $6.533\times10^9$ $N.m^{-3}$ | 0.25% |

Our analytical results fit with FEA results confirming that the nonlinear beam behavior is mainly due to beam elongation. Finally, if the beam width *b* is fixed by the fabrication process (thickness of the wafer), beam constants $k_0$ and $k_3$ can be adjusted by adjusting beam length *L* and beam thickness *e*. This design brings strong nonlinear effects that are not easy to adjust. To add a degree of freedom that will release constraints in nonlinear behaviors and to allow an optimization of nonlinear effects, we have worked on a new spring design.

## ADJUSTMENT OF NONLINEAR EFFECTS THANKS TO AN H-SHAPED SPRING DESIGN

In order to reduce nonlinear effects and to add an adjustment parameter that will allow to optimize VEH output power, we have developed an H-shaped design for beams.

### H-shaped spring design

As presented in the previous sections, the nonlinear effect of a clamped-guided beam is mainly due to beam elongation contributions. To adjust nonlinear effects, we add a spring, mechanically in series with the beam length, in order to make the elongation easier and then to reduce the contribution of the beam elongation in counter reaction to the force F. This additional spring has to block all the rotations and two translations, and acts as a spring only in $\vec{e}_y$ direction. To achieve this, we choose an H structure with a central part blocked in rotation by four beams that can move only in flexion and characterized by a global spring stiffness $k_{parr}$. The spring structure is presented in figure 7. The clamped boundary is now fixed to 4 clamped-guided beams in flexion, referred as 'holding beams'. These beams size $l_1 \times a \times b$ or $l_3 \times a \times b$. Their goal is to soften the stiffness of the main beam in traction and have an effect only on the nonlinear behavior (order 3).

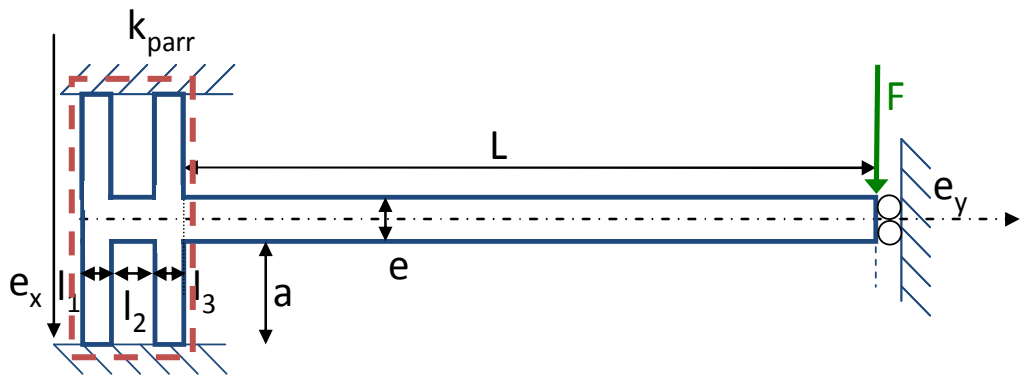

**Figure 7. H-shaped spring design**

We propose in the following part to introduce this additional spring effect in our analytical relation between the force *F* and the associated displacement *x*.

**Adjustable nonlinear effect**

As presented in section "clamped guided beam and nonlinear effects", nonlinear effects are essentially due to traction effects and elongation phenomena. As a consequence, at the first order (linear), $k_{parr}$ does not have any influence on the system as there is no elongation of the main beam. But, when elongation phenomena on the main beam starts to appear (order 3), the additional H spring of constant $k_{parr}$ acts in series with the beam stiffness in its length direction $k_{trac}$ as shown in figure 8 (a, b, c). An equivalent model of the H-shaped spring is presented figure 8(b) (linear) and 8(c) (nonlinear).

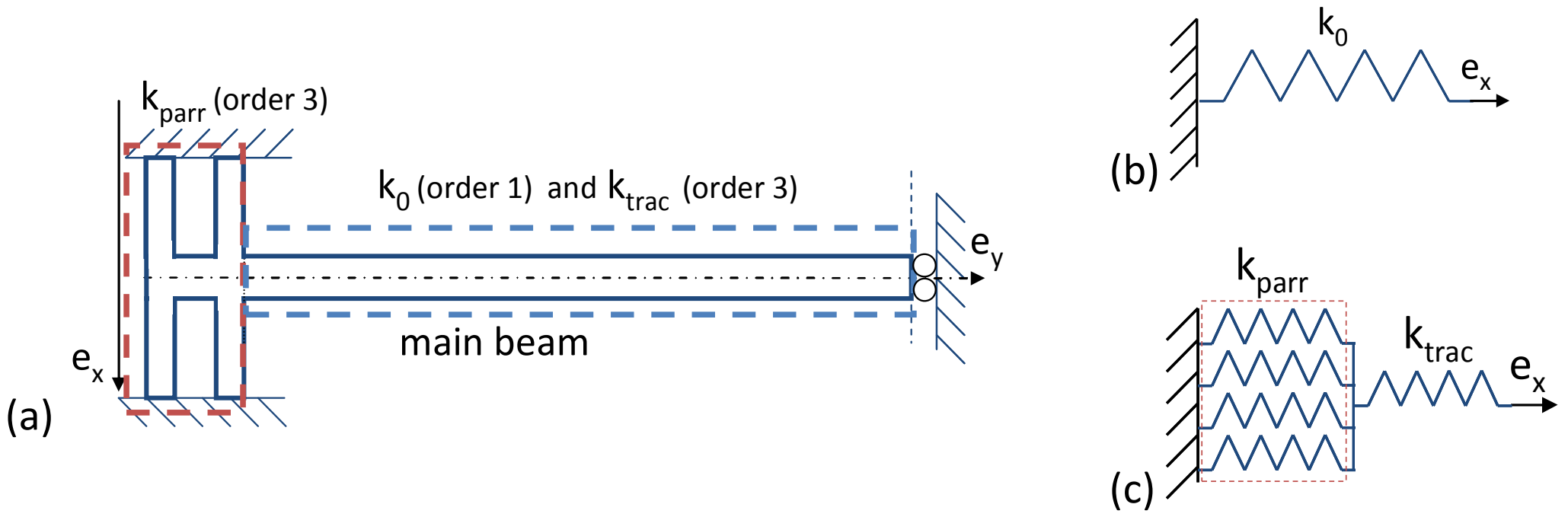

**Figure 8. (a) Model of the new spring and equivalent behavior in (b) linear domain and (c) nonlinear domain**

From structural mechanics, we can express the stiffness of $k_{parr}$ made of 4 clamped-guided beams in parallel ($k_{parr}$ stiffness is the sum of the 4 clamped-guided beams' stiffnesses that constitute it) and $k_{trac}$ the beam traction spring constant. We remind $k_0$, the spring constant of the main beam in its linear mode.

$$k_{parr} = \frac{Eb}{a^3}\left(2l_1^3 + 2l_3^3\right) , \; k_{trac} = \frac{Ebe}{L} , \; k_0 = \frac{Ebe^3}{L^3} \tag{14}$$

Therefore, to deduce the nonlinear coefficient $k_3$ of the new spring, we just have to replace $k_{trac}$ in equation (11) that represents traction phenomena by its new value $\left(\frac{k_{trac} k_{parr}}{k_{parr} + k_{trac}}\right)$ that corresponds to the equivalent spring stiffness of $k_{trac}$ and $k_{parr}$ in series. In order to simplify $k_3$ expression, we introduce a new parameter $\alpha$ which corresponds to the ratio between the new and the old traction stiffness of the beam. $k_3$ is then simply expressed as a function of $k_0$.

$$k_3 = \frac{1}{2L^2}\left(\frac{k_{trac}k_{parr}}{k_{parr}+k_{trac}}\right) = k_0\,\frac{\alpha}{2e^2} \tag{15}$$

Finally, by combining flexion (linear) and traction (nonlinear) effects, the new spring force-displacement relation becomes:

$$F_{K\alpha}(x) = k_0 x\left(1+\frac{\alpha}{2}\left(\frac{x}{e}\right)^2\right) \text{with } \alpha \in [0,1] \tag{16}$$

$\alpha$ acts directly on the nonlinear effect without having any impact on the linear behavior. $\alpha$=1 corresponds to the standard clamped-guided beam while $\alpha$=0 corresponds to a “perfect” linear spring.

Once more, to validate this analytical force-displacement relation, we develop a FEA model using Comsol® Multiphysics.

**Validation using finite element analyses and limits of the model**

The same protocol developed in the case of the simple clamped-guided beam is applied in this study. An example of FEA results is given in figure 9. Figure 9(a) shows the deformation and Von Mises stresses for the whole beam and for a given x. Figures 9(b) to 9(f) present the deformation of the beam for increasing *x*. As expected the 4 holding beams acts in flexion along $e_y$ axis when the main beam is in flexion along $e_x$ axis.

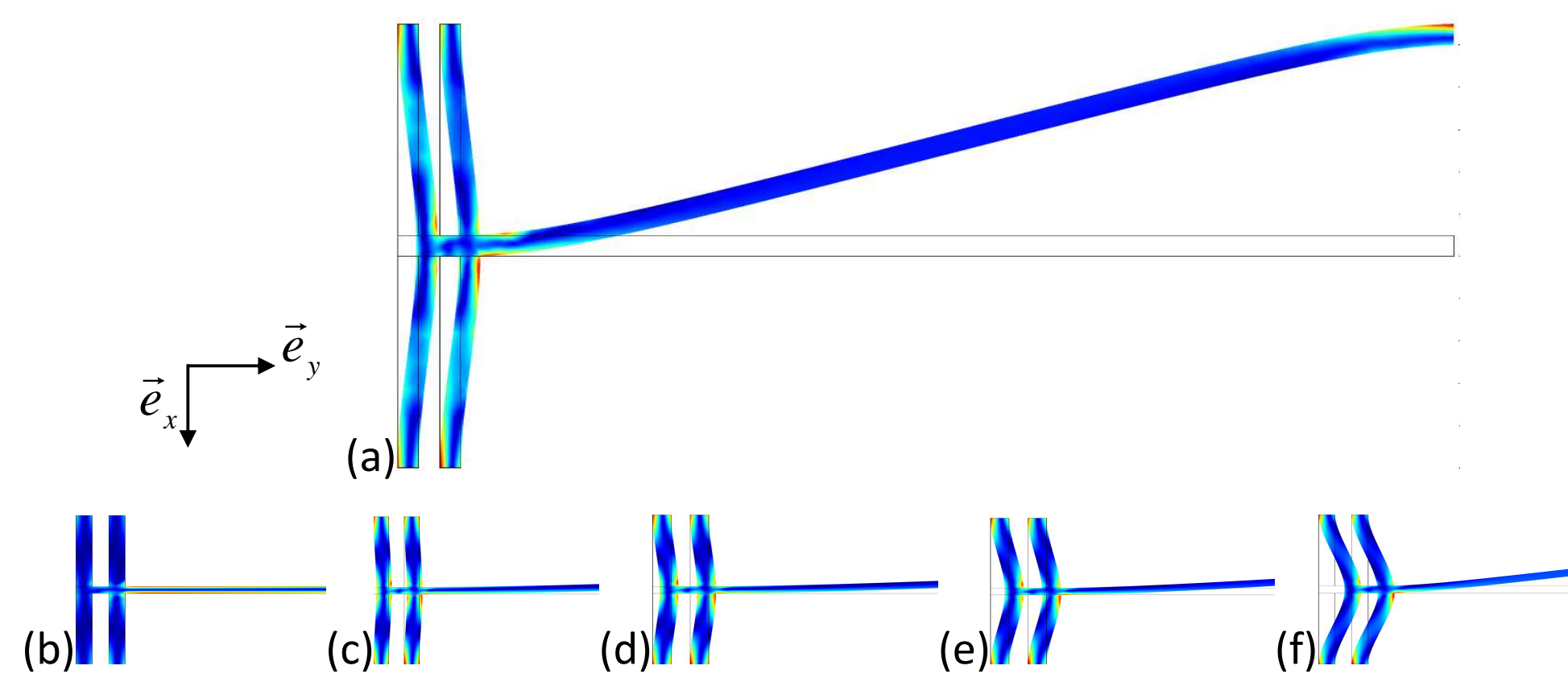

**Figure 9. Finite element analyses results. Deformation and Von Mises stresses for different x. a) complete beam and b)-to-f) zoom on the holding beams for increasing x.**

In table 2, FEA results are compared to analytical results. *L*, *b*, *e*, $l_1$, $l_2$ and $l_3$ are kept constant and respectively equals to (1cm, 1mm, 100µm, 100µm, 100µm, 100µm); *a* varies from 100µm to 1mm. It shows that the analytical model is valid and in agreement with FEA results. However, the smaller α, the larger the difference between FEA and analytical results. We think that this is due to some rotation effects that appear when holding beams are too soft compared to the main beam.

**Table 2. Theoretical and FEA values of $k_0$ and $k_3$**

| a (m) | $k_0$ (FEA) ($N.m^{-1}$) | $k_3$ (FEA) ($N.m^{-3}$) | α (FEA) | $k_{trac}$ ($N.m^{-1}$) | $k_{parr}$ ($N.m^{-1}$) | $k_0$ (theory) ($N.m^{-1}$) | $k_3$ (theory) ($N.m^{-3}$) | α (theory) | difference on $k_3$ (%) |
|---|---|---|---|---|---|---|---|---|---|
| $1\times10^{-4}$ | 130,7 | $6.50\times10^{9}$ | 1 | $1.31\times10^{6}$ | $5.24\times10^{8}$ | 131 | $6.53\times10^{9}$ | 1 | 1% |
| $2\times10^{-4}$ | 130,4 | $6.30\times10^{9}$ | 0.96 | $1.31\times10^{6}$ | $6.55\times10^{7}$ | 131 | $6.42\times10^{9}$ | 0.98 | 2% |
| $3\times10^{-4}$ | 130,0 | $5.90\times10^{9}$ | 0.9 | $1.31\times10^{6}$ | $1.94\times10^{7}$ | 131 | $6.14\times10^{9}$ | 0.94 | 4% |
| $4\times10^{-4}$ | 130,0 | $5.40\times10^{9}$ | 0.84 | $1.31\times10^{6}$ | $8.19\times10^{6}$ | 131 | $5.65\times10^{9}$ | 0.86 | 4% |
| $5\times10^{-4}$ | 129,8 | $4.90\times10^{9}$ | 0.76 | $1.31\times10^{6}$ | $4.19\times10^{6}$ | 131 | $4.99\times10^{9}$ | 0.76 | 2% |
| $6\times10^{-4}$ | 129,5 | $4.20\times10^{9}$ | 0.64 | $1.31\times10^{6}$ | $2.43\times10^{6}$ | 131 | $4.25\times10^{9}$ | 0.64 | 1% |
| $7\times10^{-4}$ | 129,2 | $3.60\times10^{9}$ | 0.56 | $1.31\times10^{6}$ | $1.53\times10^{6}$ | 131 | $3.53\times10^{9}$ | 0.54 | 2% |
| $8\times10^{-4}$ | 129,0 | $3.00\times10^{9}$ | 0.46 | $1.31\times10^{6}$ | $1.02\times10^{6}$ | 131 | $2.87\times10^{9}$ | 0.44 | 4% |
| $9\times10^{-4}$ | 128,9 | $2.50\times10^{9}$ | 0.36 | $1.31\times10^{6}$ | $7.19\times10^{5}$ | 131 | $2.32\times10^{9}$ | 0.36 | 8% |
| $10\times10^{-4}$ | 128,7 | $2.00\times10^{9}$ | 0.32 | $1.31\times10^{6}$ | $5.24\times10^{5}$ | 131 | $1.87\times10^{9}$ | 0.28 | 7% |

We have developed the analytical expression of $k_0$ and $k_3$ on our H-shaped spring design, and validated these results using FEA. As expected, the value of $k_0$ is unchanged while $k_3$ decreases with α (the spring is softened). We focus now on the effects and benefits of this new type of nonlinear springs on VEH.

## EFFECTS AND BENEFITS OF H-SHAPED NONLINEAR SPRINGS ON VELOCITY-DAMPED VIBRATION ENERGY HARVESTERS

The use of nonlinear springs modifies VEH' behaviors; in fact the harvester structure does not have a specific natural frequency; it now depends on the relative displacement amplitude between the seismic mass and the frame. We propose, in the following sub-parts, to estimate the equivalent structure natural frequency in function of the relative displacement amplitude and then to analyze the effect on VEH.

### Equivalent natural frequency

The first effect of using nonlinear springs is a natural frequency changing with the relative displacement amplitude *x(t)*. To calculate the structure equivalent natural frequency ($f_{eq}$), we use again the theory of energy conservation, neglecting electrical and mechanical dampings. Therefore, the sum of mechanical kinetic and potential energy is constant through time, and equal to the mechanical energy at $t$=0. We choose $t$=0 as a time when the mass is at its maximum displacement $x_{max}$ and has zero velocity. This equation corresponds to an unforced Duffing oscillator (Duffing, 1918; Holmes, 1980; Ueda, 1980).

$$(\dot{x})^2 + \omega_0^{\,2} x^2 + \frac{\alpha}{2e^2}\omega_0^{\,2} x^4 = constant \Rightarrow \dot{x} = \pm\omega_0\sqrt{\left(x_{max}^{\,2} - x^2\right) + \frac{\alpha}{2e^2}\left(x_{max}^{\,4} - x^4\right)} \qquad (17)$$

Equation (17) is a differential equation that cannot be solved without using a numerical solver. Nevertheless, it is possible to find $f_{eq}$ from the relative speed $\dot{x}$ without solving the differential equation.

As there is no loss of energy, the system will oscillate between $x_{max}$ and $-x_{max}$ with a frequency $f_{eq}$. Therefore, by determining the time needed to go from $x_{max}$ to *0*, we can deduce the frequency $f_{eq}$ (figure 10).

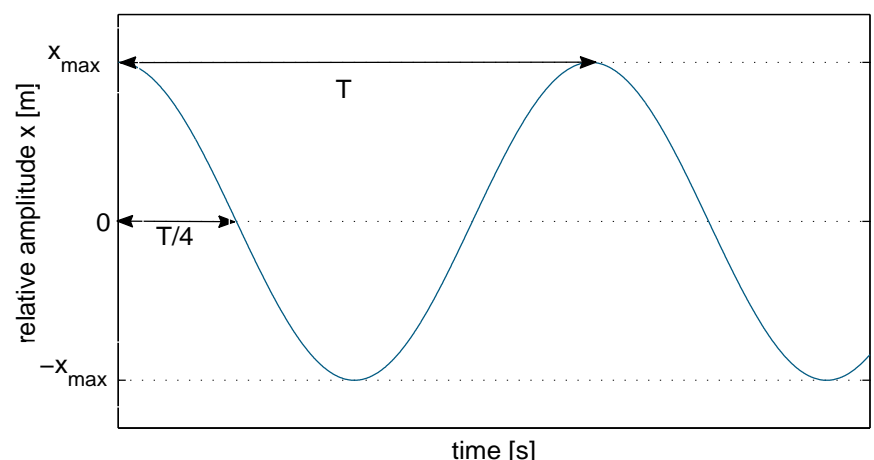

**Figure 10. Relative position (x) as a function of time (t) for non-damped oscillators**

The time needed to go from $x_{max}$ to $0$ is $T/4$ if $T$ is the oscillations period (figure 10). This time can be calculated by integrating the inverse of relative speed $\dot{x}$ as follows:

$$dt = \frac{dx}{\dot{x}} \Rightarrow T = \frac{4}{\omega_0} \int_{x=0}^{x=x_{max}} \frac{dx}{\sqrt{\left(x_{max}{}^2 - x^2\right) + \frac{\alpha}{2e^2}\left(x_{max}{}^4 - x^4\right)}} = \frac{4}{\omega_0} \frac{K\left(-\frac{\lambda}{1+\lambda}\right)}{\sqrt{1+\lambda}} \tag{18}$$

Where $\lambda = \frac{\alpha x_{max}^2}{2e^2}$ and $K(z)$ is the complete elliptic integral of the first kind defined by: $K(z) = \int_0^1 \frac{dt}{\sqrt{(1-t^2)(1-z^2t^2)}}$

If $\alpha$=0, then $\lambda$=0. Since $K(0)=\pi/2$, we found the well-known relation $T=2\pi/\omega_0$.

From (18), we can deduce $f_{eq}$ as a function of $x_{max}$ since $f_{eq}=1/T$. $f_{eq}$ is function of the relative displacement amplitude, the higher the displacement amplitude, the higher the natural frequency becomes.

**Evolution of equivalent natural frequency**

Nonlinear springs' effects on the VEH natural frequency becomes significant when the relative displacement amplitude is higher than the beam thickness $x_{max}>e$, as expected. Figure 11 shows the normalized equivalent natural frequency $f_{eq}/f_0$ as a function of the normalized maximum relative displacement $x_{max}/e$ with $e$ the beam thickness.

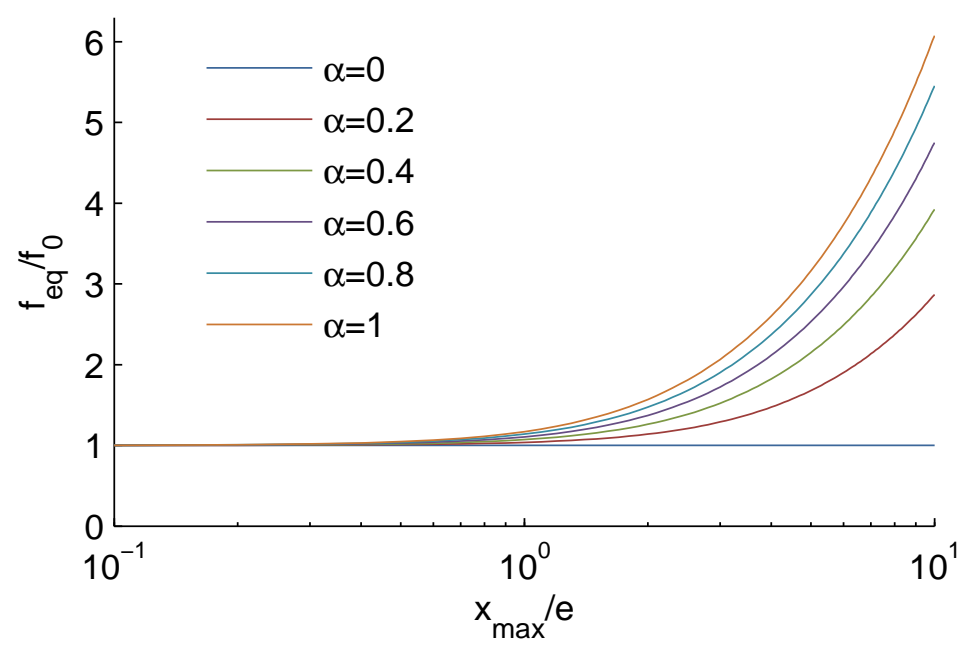

**Figure 11. Variation of the normalized natural frequency $f_{eq}/f_0$ of the energy harvester as a function of the normalized displacement amplitude $x_{max}/e$ ($f_0$=100Hz, e=100µm)**

For a clamped-guided beam ($\alpha$=1), when $x_{max}=10\times e$, $f_{eq}$ is multiplied by 6 compared to the linear natural frequency $f_0$.

Uses of nonlinear springs have a significant impact on the VEH natural frequency and we can suppose that it will also significantly impact the harvester capability to extract vibration energy and turn it into electrical output power. We propose in the following part to study the nonlinear effects on the VEH output power in function of the input vibration frequency and amplitude.

## Effects on vibration energy harvesting

In order to solve differential equations and to determine VEH's behavior to "ambient" vibrations, Simulink models have been developed for linear and non-linear VEH. These models allow to get mobile masses' relative movements and their derivatives as well as VEH output powers in both cases (figures 12(a) and 12(b)).

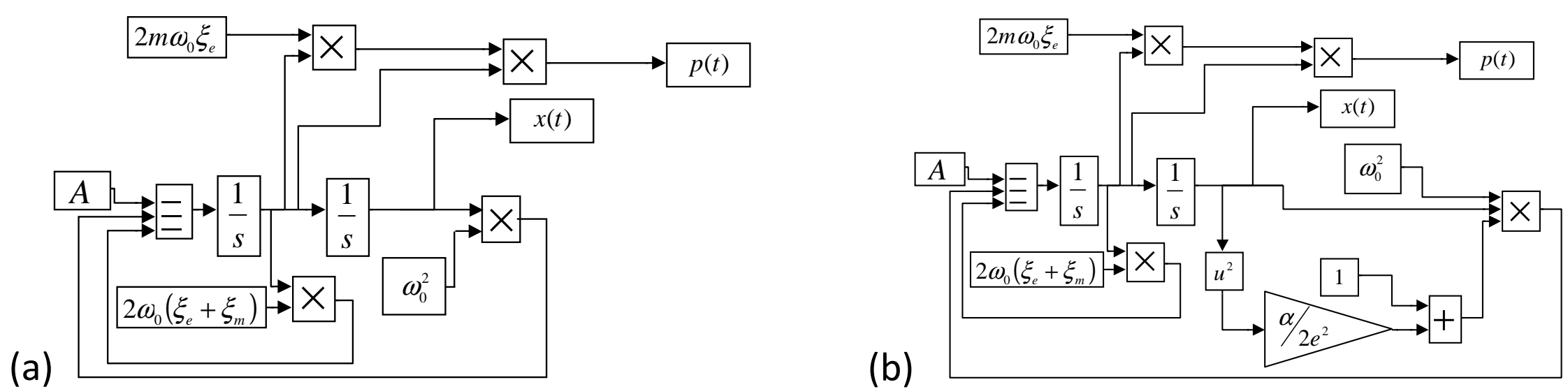

**Figure 12. Simulink model of VEH (a) linear and (b) nonlinear**

In figure 13 is presented VEH's relative displacement and output power for different values of α, for an acceleration level *A* of 1 m.s$^{-2}$ and for varying frequencies. We precise again that *α=0* corresponds to a linear energy harvester (no beam traction effect) and *α=1* is a full clamped-guided beam. In all cases, $f_0$=100Hz, the mechanical quality factor $Q = \frac{1}{2\xi_m}$ is chosen equal to 100 and the electrical damping is maintained equal to the mechanical damping $\xi_e=\xi_m$.

$X_{linear}$ and $P_{linear}$ are respectively the maximum relative displacement and the output power (for 1g of mobile mass) of the linear VEH (*α*=0). $X_{up}$ and $P_{up}$ are the maximum relative displacement and the output power (for 1g of mobile mass) for a nonlinear VEH and for an increasing vibration frequency, while $X_{down}$ and $P_{down}$ are obtained for a decreasing vibration frequency.

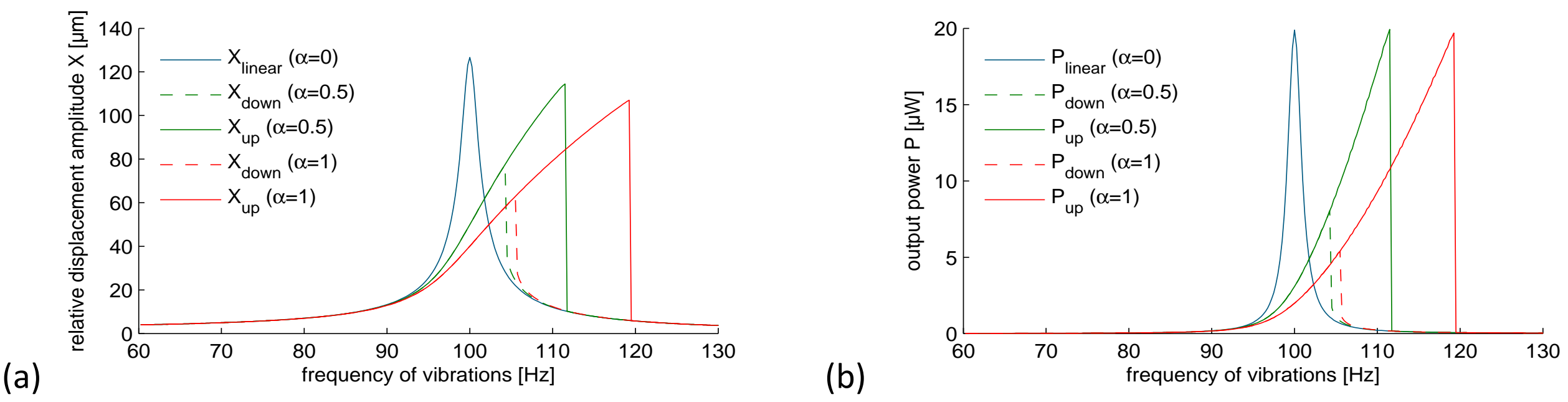

**Figure 13. (a) Relative displacement and (b) output power of a nonlinear energy harvester as a function of α (A=1 m.s$^{-2}$) for e=100µm**

As expected, for linear VEH, the maximum of output power is reached when the vibration frequency *f* equals the natural frequency $f_0$ and the operating frequency bandwidth is quite low.

As for nonlinear VEH, the behavior is not the same with increasing frequencies ($X_{up}$ and $P_{up}$) and decreasing frequencies ($X_{down}$ and $P_{down}$). This hysteresis is due to the fact that both solutions are viable for some frequencies. If the relative displacement amplitude starts from a low value and if the vibration frequency is higher than the linear natural frequency, the system is not able to reach a relative amplitude that brings to an equivalent resonant frequency close to the vibration frequency: the relative displacement stays low. But, if the vibration frequency increases slowly, the relative amplitude is close to the amplitude that leads to an equivalent resonant frequency close to the input frequency: the relative displacement is amplified by the resonance effect, maintaining an amplitude that keeps the equivalent resonant frequency close to the input frequency.

We cannot ensure to stay on the upper solution; that is the major limitation of nonlinear VEH. We nevertheless notice that it is possible to follow ambient vibration frequencies' shifts with nonlinear VEH. It is also important to note that this phenomenon is passive. Moreover, the maximum output power of VEH can reach the linear maximum output power for an increasing frequency.

We have also studied the behavior of the VEH for different ambient vibrations amplitudes, while $\alpha$ is kept constant and equal to 1:

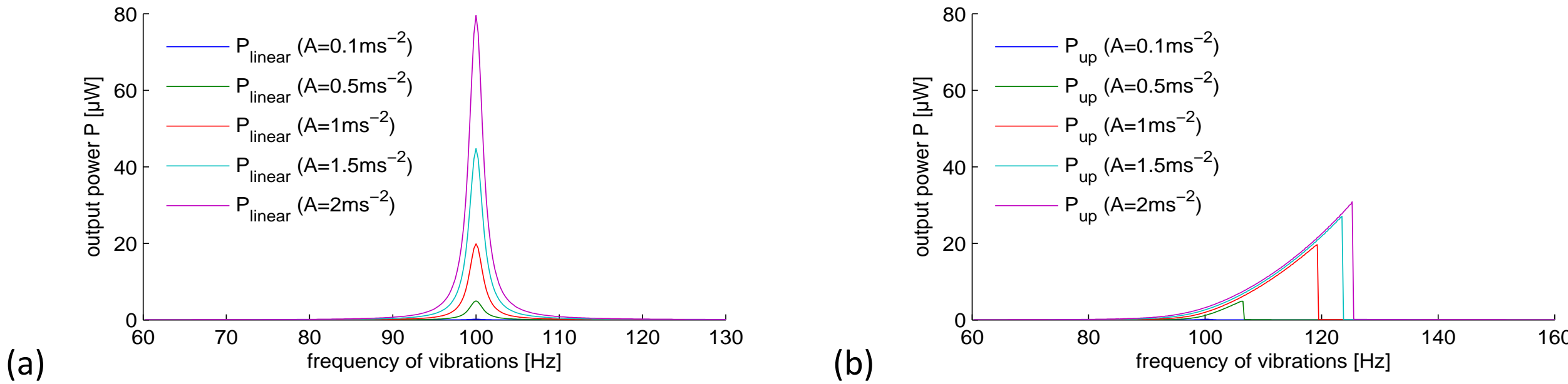

**Figure 14. Output powers of (a) a linear energy harvester and (b) a nonlinear energy harvester as a function of A (α=1)**

Figure 14 shows another great interest of nonlinear VEH: when the amplitude increases, the "upper state" is longer. In linear VEH, the working frequency bandwidth is constant whatever the acceleration. However, the maximum output power of a VEH is generally higher with linear springs. Therefore, there is a compromise to make between the maximum output power and the frequency band where the fundamental vibration frequency can be harvested; the interest of nonlinear springs compared to linear springs is not obvious and will probably depends on the vibration source.

The next section presents some examples of ambient vibration sources and shows the interest (or not) of using nonlinear springs.

## LINEAR AND NONLINEAR SPRINGS IN REAL VIBRATING ENVIRONMENTS

In this section, the previous results are applied to real vibrating environments. Actually, it is hard to know if it is better to use linear or nonlinear springs for VEH as, for a constant frequency of excitation, linear VEH are better than nonlinear ones in terms of output power. After a presentation of vibration sources, we expose the maximum power that can be harvested using linear and nonlinear devices thanks to an optimization process.

### Application to a real environment: energy harvesting from a car engine

We present in this section a first example of an ambient vibration source and the interest of using nonlinear springs. The vibration source is a car engine at 3000rpm. The accelerometer has directly been placed on the engine to measure the temporal acceleration of vibrations. Figures 15 (a, b, c) present the temporal and spectral acceleration of these 'ambient' vibrations.

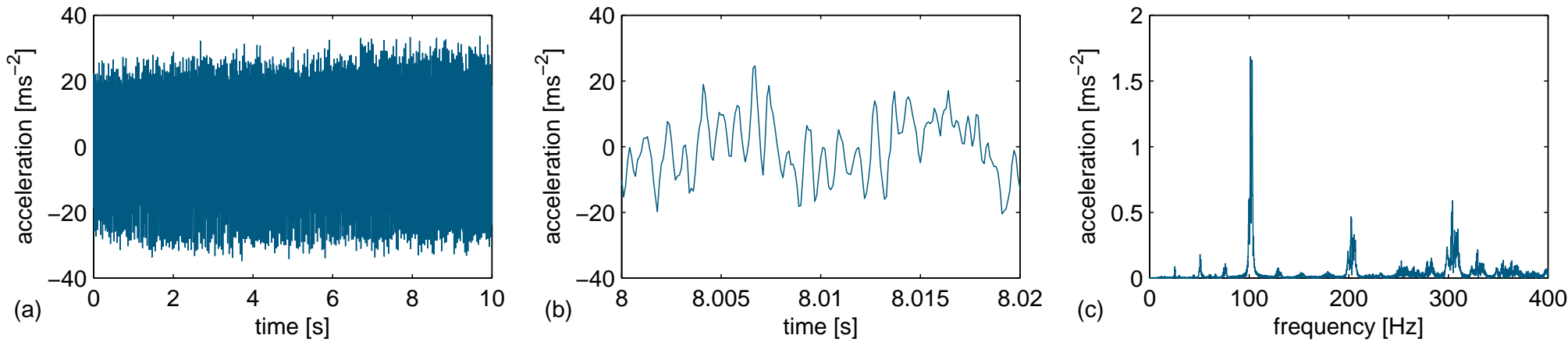

**Figure 15. Vibrations on a car: (a) temporal (b) zoom between 8 and 8.02s and (c) spectrum**

The next study compares the output power that can be harvested with optimized linear and nonlinear VEH.

**Linear vs nonlinear springs on a car engine**

To compare the effects of linear and nonlinear springs on VEH, we have taken the same ambient vibration source "car engine at 3000rpm" and use it as an input data in the two Simulink models (linear and nonlinear). An optimization process has been developed and implemented in Matlab using *fminsearch* function, which is aimed at maximizing the VEH output power and takes ($f_0$, $\xi_e$) as input parameters for the linear system and ($f_0$, $\xi_e$, $\beta = \alpha / 2e^2$ ) for the nonlinear VEH. The mechanical quality factor $Q$ is 100. In each case, while $p(t)$ is the instantaneous output power, the "output power" P is the mean output power of the VEH, computed on the whole sample duration (10s). The power is given per gram of mobile mass.

On one hand, the maximum output power $P$ reached by the linear VEH is 483.57µW/g, obtained for ($f_0$, $\xi_e$)=(102.2 Hz, 0.0089). Figure 16(a, b) show the instantaneous relative displacement and figure 16(c, d) the output power for these optimized parameters:

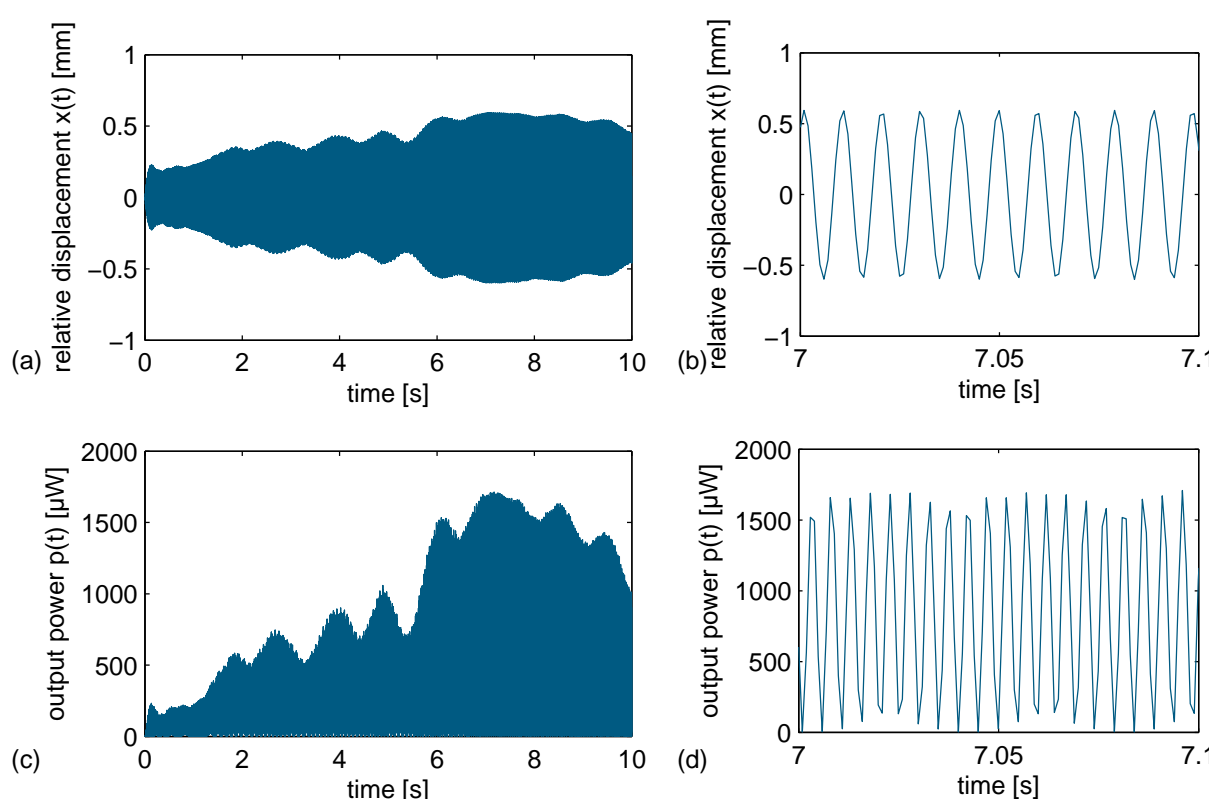

**Figure 16. (a) Relative displacement and (b) zoom; (c) output power of the linear energy harvester for m=1g and (d) zoom.**

If the acceleration level seems to be constant on all the sample duration (figure 15(a)), the relative displacement filtered by the linear system and the associated instantaneous output power changes significantly.

On the other hand, the maximum output power $P$ reached by the nonlinear VEH is 717.75µW/g and obtained for ($f_0$, $\xi_e$, β)=(98.77 Hz, 0.00539, $1.56 \times 10^5$ $m^{-2}$); this represents 48% more output power than linear VEH. Figure 17(a, b) show the instantaneous relative displacement and figure 17(c, d) the output power for these optimized parameters:

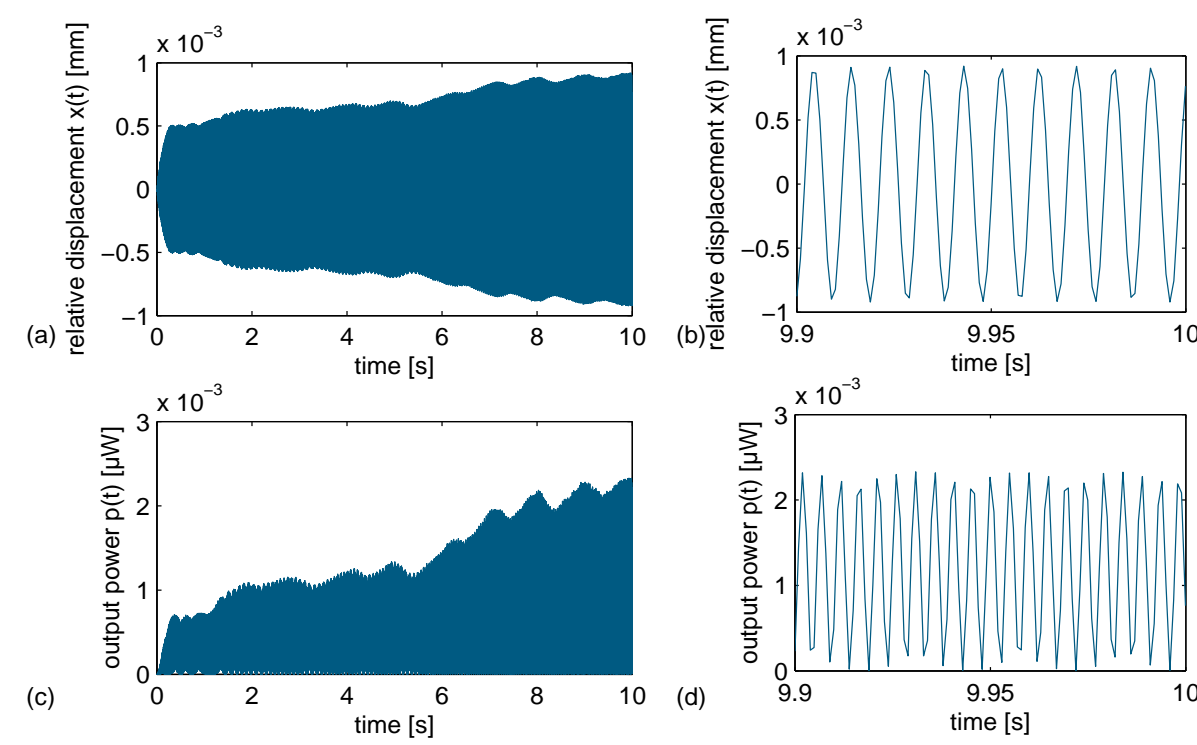

**Figure 17. (a) Relative displacement and (b) zoom; (c) output power of the nonlinear energy harvester for m=1g and (d) zoom**

Therefore, we have proven the interest of nonlinear springs to improve the output power of VEH placed on a car engine. For the vibration source considered, nonlinear springs allow to increase by 48% the output power of VEH compared to linear springs.

### Nonlinear springs in other vibrating environments

We present in this paragraph the interest (or not) of nonlinear springs for 3 other ambient vibrations sources. We have chosen a staircase with someone going down, a drill and the same car engine as previously but with a 2000rpm rotation speed.

**Table 3. 3 other sources of vibration**

| Source of vibrations | Acceleration spectrum | Optimum for linear devices | Optimum for nonlinear devices |
|---|---|---|---|
| car engine at 3000rpm – previous paragraph | | $f_0$=102.2 Hz<br>$\xi_e$=0.0089<br>$P$=483.57μW | $f_0$=98.77 Hz<br>$\xi_e$=0.00539<br>β=1.56×10$^5$ m$^{-2}$<br>$P$=717.75 μW |
| staircase with someone going down | | $f_0$=11.7 Hz<br>$\xi_e$=0.0093<br>$P$=80μW | $f_0$=7.93 Hz<br>$\xi_e$=0.0165<br>β=5.144×10$^3$ m$^{-2}$<br>$P$=100 μW |
| drill | | $f_0$=99.99 Hz<br>$\xi_e$=0.0051<br>$P$=70.6μW | $f_0$=99 Hz<br>$\xi_e$=0.052<br>β=4.74×10$^5$ m$^{-2}$<br>$P$=71.1 μW |
| car engine at 2000rpm | | $f_0$=131.84 Hz<br>$\xi_e$=0.0065<br>$P$=26.92 μW | $f_0$=131.84 Hz<br>$\xi_e$=0.0065<br>β=0 m$^{-2}$<br>$P$=26.92 μW |

This study proves the interest of nonlinear VEH for spread frequency spectrum vibrations as in staircase. Nonlinear springs have a lower interest in the other cases: the drill and the car engine at stabilized 2000rpm, as the vibration energy is concentrated on a very narrow frequency bandwidth.

## PERSPECTIVE – APPLICATION TO AN ELECTROMAGNETIC ENERGY HARVESTER

In this section, we present a possible design for an electromagnetic VEH using the previous nonlinear springs in order to prove the feasibility of the concept.

### Electromagnetic energy harvester using the H-shaped spring design

It is possible to imagine a simple VEH made of a mass *m* maintained thanks to two beams in steel (E=200GPa). A diagram of a possible prototype is presented in figure 18. The mass moves in a coil when the device is subjected to ambient vibrations. Electric power is generated due to the relative movement of a magnet into a coil (Lenz's law) and circulates through the load.

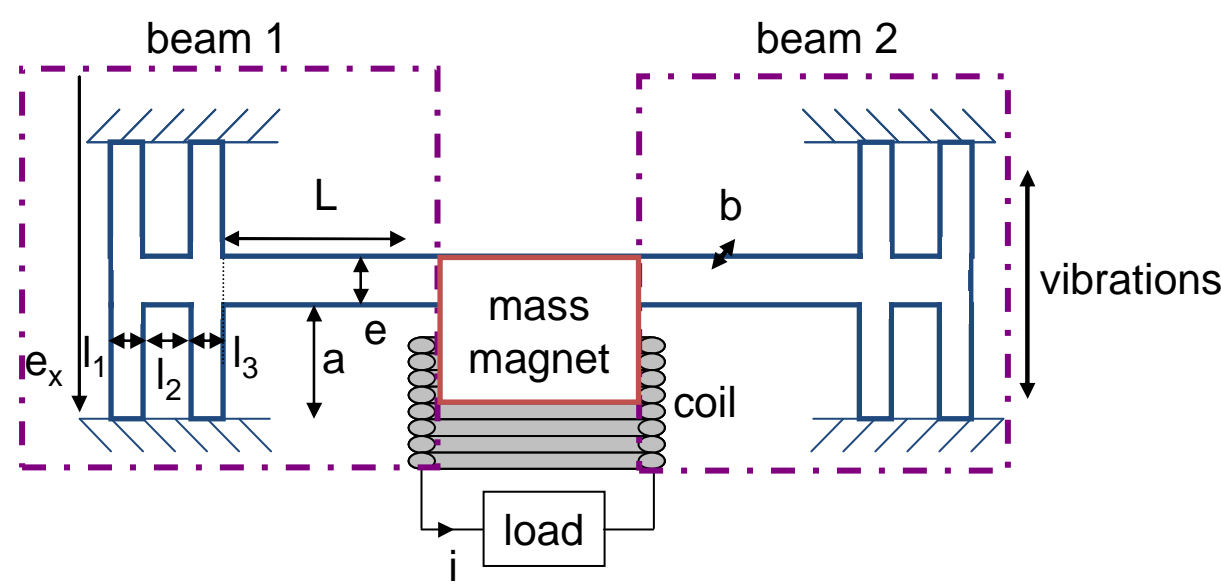

**Figure 18. Nonlinear piezoelectric energy harvester**

As the two beams act in parallel, the global stiffness of the VEH $k_{EH}(x)$ is:

$$k_{EH}(x) = 2k(x) = 2k_0 + 2k_0 \underbrace{\left(\frac{\alpha}{2e^2}\right)}_{\beta} x^2 \tag{19}$$

### Example of mechanical sizing for the nonlinear electromagnetic energy harvester

We propose here to optimize the nonlinear electromagnetic VEH for the vibration source presented in the previous paragraphs: the engine of a car at 3000rpm.

We present two designs of viable VEH to respect ($f_0$, β)=(98.77 Hz, $1.56\times10^5$ m$^{-2}$). The first has a mobile mass of 1g and theoretically allows to harvest 717.75µW and the second can harvest 7.17mW with a mobile mass of 10g.

**Table 4. Two viable dimensioning for a car engine at 3000tpm**

| parameter | Designation | m=1g | m=10g |
|---|---|---|---|
| $f_0$ | natural frequency of the energy harvester | 98.77Hz | 98.77Hz |
| $k_{0EH}$ | total linear spring stiffness of the energy harvester | 385.15 N/m | 3851.57 N/m |
| $k_0$ | linear spring stiffness of one beam | 192.57 N/m | 1925.78 N/m |
| b | width of the beam | 2cm | 2cm |
| L | length of the beam | 5.5cm | 6.38cm |
| e | thickness of the beam | 200µm | 500µm |
| $l_1$, $l_3$ | width of the holding beams | 100µm | 500µm |
| a | length of the holding beams | 0.44cm | 0.91cm |
| $k_{parr}$ | linear spring constant of the holding beams | $1.88\times10^5$ N/m | $2.65\times10^6$ N/m |
| $k_{trac}$ | spring constant of the beam in traction | $1.46\times10^7$ N/m | $3.14\times10^7$ N/m |
| $k_{3th}$ | theoretical spring constant of the beam – order 3 | $3\times10^7$ N/m³ | $3\times10^8$ N/m³ |
| $k_{3FEA}$ | FEA spring constant of the beam – order 3 | $4\times10^7$ N/m³ | $3.6\times10^8$ N/m³ |
| relative difference between $k_{3th}$ et $k_{3FEA}$ | | 28% | 18% |
| $P_{th}$ | theoretical output power | 717.75µW | 7.17mW |

Theoretical and FEA results correspond fairly well (28% of difference on $k_3$ in the worst case). As α for each beam is small ($6.2\times10^{-3}$ for the 1g case and $3.9\times10^{-2}$ for the 10g case) and according to results presented in table 2, this difference was more ore less expected. Once again, we think that those results are due to rotation effects that appear when holding beams are too soft compared to the main beam. Obviously, the different values can be then adjusted using FEA for more precision but the formula gives nevertheless a good order of magnitude for the different dimensions even with small α.

New simulations have proven that to reach $\beta=1.56\times10^5$ $m^{-2}$ in both cases, *a* should be increased to 0.5cm (instead of 0.44cm) in the 1g case and to 1cm (instead of 0.91cm) in the 10g case.

**Equivalent electrical damping**

We have proven that, to maximize VEH output power, it is necessary to apply an electrical damping rate $\xi_e$=0.00539. The question is now to know if it is possible to reach this value using the sized previous electromagnetic structure.

From (Roundy, 2006), the equivalent electrical damping of an electromagnetic converter can be expressed as:

$$\xi_e = \frac{(NlB)^2}{2m\omega R} \tag{20}$$

Where N, is the number of turns of the generator coil, L the side length of the coil (supposed to be a square) and B the flux density to which the system is subjected. This formula is valid as soon as the electrical time constant is small with regards to the mechanical one.

As a consequence, with a good value for *R*, an electrical damping $\xi_e$=0.00539 can be easily reached (e.g. for N=10, L=1cm, B=1T). This concludes the feasibility of our H-shaped design and the interest of nonlinear springs for vibration energy harvesting to increase, in some cases, VEH output power.

**CONCLUSIONS & PERSPECTIVES**

We have developed models of clamped-guided beams, taking nonlinear effects at the order 3 into account and based on the theory of energy conservation. We have also designed new springs allowing to adjust these nonlinear effects, while the spring-force relation at the order 1 stays the same. These models have been applied to a real source of vibrations, the engine of a car, and the results proved the interest of nonlinear springs to increase the output power of resonant energy harvesters. We finally present the mechanical design of an electromagnetic energy harvester sized according to the results of the optimization for the car engine vibrations at 3000tpm; it proves the theoretical feasibility of such devices.

**REFERENCES**

Ando, B., S. Baglio, C. Trigona, N. Dumas, L. Latorre and P. Nouet. 2010. “Nonlinear mechanism in MEMS devices for energy harvesting applications,” Journal of Micromechanics and Microengineering.

Anton, S.R. and H. A. Sodano. 2007. “A review of power harvesting using piezoelectric materials (2003–2006),” Smart Materials and Structures, 16.

Amri, M., P. Basset, F. Cottone, D. Galayko, F. Najar and T. Bourouina. 2011. “Novel nonlinear spring design for wideband vibration energy harvesters,” Proc. PowerMEMS

Beeby, S.P., M.J. Tudor and N.M. White. 2006. “Energy harvesting vibration sources for microsystems applications,” Measurement science and Technology, 17.

Boisseau, S., G. Despesse and A. Sylvestre. 2010. “Optimization of an electret-based energy harvester,” Smart Material and Structures, 19.

Boisseau, S., G. Despesse and A. Sylvestre. 2011. “Cantilever-based electret energy harvesters,” Smart Material and Structures, 20.

Cook-Chennault, K.A., N. Thambi and A. M. Sastry. 2008. “Powering MEMS portable devices—a review of non-regenerative and regenerative power supply systems with special emphasis on piezoelectric energy harvesting systems,” Smart Materials and Structures, 17.

Cottone, F., H. Vocca and L. Gammaitoni. 2009. “Nonlinear Energy Harvesting,” Physical Review Letters, 102.

Despesse, G. 2005. “Etude des phénomènes physiques utilisables pour alimenter en énergie électrique des micro-systèmes communicants,” Thesis, INPG, France.

Duffing, G. 1918. ”Erzwungene Schwingungen bei Veränderlicher Eigenfrequenz.,” F. Vieweg u. Sohn, Braunschweig.

Erturk, A. and D.J. Inman. 2008. “Issues in Mathematical Modeling of Piezoelectric Energy Harvesters,” Smart Materials and Structures, 17, 065016.

Ferrari, M., V. Ferrari, M. Guizzetti, B. Ando, S. Baglio and C. Trigona. 2009. “Improved Energy Harvesting from Wideband Vibrations by Nonlinear Piezoelectric Converters,” Proc. Eurosensors’09:1203-1206.

Holmes, P. and D. Rand. 1980. “Phase portraits and bifurcations of the non-linear oscillator: $\ddot{x}+(\alpha+\gamma x^2)\dot{x}+\beta x+\delta x^3=0$,” International Journal of Non-linear Mechanics, 15: 449-458.

Kaajakari, K., T. Mattila, A. Oja and H. Seppa. 2004. “Nonlinear limits for single-crystal silicon microresonators,” Journal of Microelectromechanical systems, 13: 715-724.

Landau, L.D. and E.M. Lifshitz. 1999. “Mechanics, 3rd edition,” Butterworth-Heinemann, Oxford.

Legtenberg, R., A. W. Groeneveld and M. Elwenspoek. 1996. “Comb-drive actuators for large displacements,” Journal of Micromechanics and Microengineering, 6: 320-329.

Lobontiu, N., and E. Garcia. 2005. “Mechanics of microelectromechanical systems,” Springer.

Mann, B.P. and B.A. Owens. 2010. “Investigations of a nonlinear energy harvester with a bistable potential well,” Journal of Sound and Vibrations, 329: 1215-1226.

Marzencki, M., M. Defosseux and S. Basrour. 2009. “MEMS Vibration Energy Harvesting Devices With Passive Resonance Frequency Adaptation Capability,” Journal of Microelectromechanical Systems, 18: 1444-1453.

Miki, D., M. Honzumi, Y. Suzuki and N. Kasagi. 2010. “Large-Amplitude mems electret generator with nonlinear springs,” Proc. MEMS’10: 176-179.

Nguyen, D.S. and E. Halvorsen. 2010. “Analysis of vibration energy harvesters utilizing a variety of nonlinear springs,” Proc. PowerMEMS’10: 331-334.

Quinn, D., A. Triplett, L. Bergman and A. Vakakis. 2011. “Comparing Linear and Essentially Nonlinear Vibration-Based Energy Harvesting,” Journal of vibration and acoustics, 133.

Roundy, S., P. K. Wright and J. Rabaey. 2003. “A study of low level vibrations as a power source for wireless sensor nodes,” Computer Communications, 26: 1131-1143.

Saadon, S. and O. Sidek. 2011. “A review of vibration-based MEMS piezoelectric energy harvesters,” Energy Conversion and Management, 52: 500-504.

Stanton, S.C. and B.P. Mann. 2010. “Engaging nonlinearity for enhanced vibratory energy harvesting,” Structural Dynamics and Materials.

Tang, L., Y. Yang and C. Soh. 2010. “Toward Broadband Vibration-based Energy Harvesting,” Journal of Intelligent Material Systems and Structures, 21: 1867-1897.

Triplett, A., and D. Quinn. 2009. “The Effect of Nonlinear Piezoelectric Coupling on Vibration-based Energy Harvesting,” Journal of Intelligent Material Systems and Structures, 16: 1959-1967.

Ueda, Y. 1980. “Explosion of strange attractors exhibited by Duffing's equation,” Nonlinear Dynamics, R.H.G. Helleman (ed.), 422-434.

Wiggins, S. 1990. "Application to the Dynamics of the Damped, Forced Duffing Oscillator," Introduction to Applied Nonlinear Dynamical Systems and Chaos. New York: Springer-Verlag.

Williams, C. B. and R. B. Yates. 1996. “Analysis of a micro-electric generator for Microsystems,” Sensors and Actuators A, 52: 8-11.